\documentclass{elsart1p}

 \usepackage{graphicx}
\usepackage{amssymb}
\begin{document}
\begin{frontmatter}
\title{High-energy hadron physics at future facilities
 }

\author{Mark Strikman}

\address{104 Davey lab, Penn State University, University Park, PA16802, U.S.A.}

\begin{abstract}We  outline  several directions for future investigations  of the three-dimensional  structure of nucleon, including multiparton correlations, color transparency, and branching processes  at hadron colliders and at hadron  factories. We also 
find evidence that pQCD regime for non-vacuum Regge trajectories sets in for $-t\ge 1\, \mbox{GeV}^2$ leading to nearly t-independent trajectories.
\end{abstract}

\begin{keyword}
parton distributions  \sep hadron structure \sep color transparency

\PACS 24.85.+p \sep 12.38.-t \sep 13.85.-t
\end{keyword}
\end{frontmatter}

% main text
\section{Introduction}
\label{}
So far major successes of QCD in describing high energy hadron 
hadron collisions  were for hard inclusive processes - collision
 of two partons. To describe these processes it is
sufficient  to know only longitudinal single parton densities.

At the same time the knowledge of  the transverse spread of partons,  
longitudinal and transverse correlations of partons, their dependence
on flavor, x  polarization of the parton,   are  necessary for
building a realistic description of the 
global structure of the final states  in pp collisions, 
 understanding  microscopic structure of  nucleon bound state.

Hard exclusive processes in $ep$ scattering allow to probe not only
the distribution of partons with respect to longitudinal momentum,
but also their spatial distribution in the transverse plane.
Examples include the hard electroproduction of light mesons 
and real photons (deeply virtual Compton scattering), as well as
the photoproduction of heavy quarkonia ($J/\psi, \psi^{\prime},\Upsilon$).
Thanks to QCD factorization theorems the amplitudes for these processes 
can be separated into a ``hard'' part, calculable in perturbative QCD, 
and ``soft'' parts characterizing the non-perturbative structure of 
the involved hadrons. The information about the nucleon is contained
in so-called generalized parton distributions (GPD's). These are functions
of the parton momentum fractions, $x$ and $x'$, as well as 
of the invariant momentum transfer to the nucleon, $t$, and thus
combine aspects of the usual parton distributions, measured
in inclusive deep--inelastic scattering, with those
of the elastic nucleon form factors. For $x = x'$, their Fourier transform 
with respect to $t$ describes the spatial distribution of partons
in the transverse plane. The GPD's thus, in a sense, 
provide us with a ``3D parton image'' of the nucleon.

Extensive data on VM production from HERA support dominance of the  pQCD dynamics. Numerical calculations including  transverse size  of  the wave function of vector meson explain key elements of high   $Q^2$   data.
In particular, they confirm prediction of \cite{Brodsky}
that t-slope of convergence of the slopes of the $\rho$-meson 
diffractive electroproduction at large $Q^2$ to the slope  of the  $J/\psi$ meson 
photo(electro) production and describe the  $Q^2$ dependence of the difference 
of the slopes \cite{FKS}.  Hence the t-dependence of $J/\psi$ photo/electro production is given (up to small corrections ) by the square of two - gluon form factor of the nucleon, which is related to the transverse distribution
 of the gluons at given x as:
\begin{equation}
F_g (x, t ) \;\; = \;\; \int d^2 \rho \; 
e^{i (\vec{\Delta}_\perp \vec{\rho})} \; F_g (x, \rho) ,
\hspace{4em}
(t = -\vec{\Delta}_\perp^2 ) .
\label{impact_def}
\end{equation}
The function $F_g (x, \rho)$ 
is normalized as 
$\int d^2 \rho \, F_g (x, \rho) = 1$. 
 Concerning the shape of the two-gluon form factor, 
it has been argued that at $x \geq 10^{-1}$ (where pion cloud contributions
are absent) the two--gluon form factor should follow the axial form factor
of the nucleon, and thus be described by the dipole parameterization
\begin{equation}
F_g (x, t) \;\; = \;\; (1 - t/m_g^2)^{-2},
\hspace{4em} m_g^2 \;\; = \;\; 1.1 \, {\rm GeV}^2 
\hspace{4em} (x \geq 10^{-1}).
\label{dipole}
\end{equation}
This form indeed describes well the $t$--dependence of the fixed--target
$J/\psi$ photoproduction experiments. 
Increase of the slope between $x\sim 0.1$ and $x\sim 0.01$ appears to be mostly due to
the scattering off the pion field \cite{Strikman:2003gz}
Based on Eq.~(\ref{dipole}), 
we have suggested in Ref.~\cite{Frankfurt:2003td} a generalized
dipole parametrization valid also at small $x$, which incorporates
the observed increase in the gluonic transverse size (as well as the
effects of DGLAP evolution) by way of an $x$-- and $Q^2$--dependent
of the dipole mass parameter, $m_g^2 (x, Q^2)$. For details, see
Refs.~\cite{Frankfurt:2003td,DIS04}.

\section{Collider options}
Exclusive processes in $ep$ scattering, however, are not the
only reactions which probe the ``3D'' parton distributions.
In fact, a lot more information about the longitudinal momentum
and transverse spatial distribution of partons, as well as 
about multiparton correlations, can be obtained from the study
of selected hard processes in (not necessarily exclusive) 
$pp$ and $pA$ scattering. Comparative studies of $ep$ and $pp/pA$ 
induced hard processes will help to improve the quantitative description 
of both classes of processes and offer many new, fascinating insights 
into the partonic structure of the nucleon. 

Turning now to $pp$ collisions, an immediate application
of the transverse spatial distribution of partons is in
the description of the impact parameter dependence of the
cross section for hard dijet production \cite{Frankfurt:2003td}.
In a $pp$ collision with c.m.\ energy $\sqrt{s}$, a hard dijet with 
transverse momentum $q_\perp$ at zero rapidity is produced 
in the collision of two partons carrying momentum fractions
$x_1, x_2$. The probability for such a parton--parton 
collision, as function of the impact parameter of the $pp$ system, $b$,
is given by the convolution of the spatial distributions of the
partons; for gluons
\begin{equation}
P_2 (b) \;\; \equiv \;\; \int d^2\rho_1 \int d^2\rho_2 \; 
\delta^{(2)} (\vec{b} - \vec{\rho}_1 + \vec{\rho}_2 )
\; F_g (x_1, \rho_1 ) \; F_g (x_2, \rho_2 ) .
\label{P_2}
\end{equation}
The scale of the parton distributions here is $q_\perp^2$.

While exclusive processes in $ep$ scattering provide in principle 
the cleanest way to access the transverse spatial distribution
of partons, there are several instances in which $pp$ scattering
is more effective. One is the study of large $x$, where the
cross sections for exclusive processes in $ep$ are small. 
In this case the global transverse distribution of matter can
be measured more directly using various reactions combining 
a soft and hard trigger, in particular in connection with 
$pA$ collisions \cite{Frankfurt:1985cv}. New opportunities 
for such studies will emerge at LHC, where the high luminosity will allow, 
for example, to compare the characteristics of $W^+$ and $W^-$ production 
at the same forward rapidities, corresponding to sufficiently high $x$. 
By studying the accompanying production of hadrons one can
learn which configurations in the nucleon have larger transverse size
--- those with a leading u--quark or with a leading d-quark. 
One suitable observable is, for example, the distribution of the 
number of events over the number of the produced soft particles. 
A larger transverse size corresponds to a larger probability 
of soft interactions, and hence to a larger probability of events 
with large multiplicity. It is interesting to note that the 
studies of the associated soft hadron multiplicity in the production 
of $W^{\pm}$ and $Z$ bosons in $\bar p p$ collisions at $\sqrt{s}=2\, TeV$ find an increase of this multiplicity 
by a factor of two as compared to generic inelastic 
events \cite{Field:2002vt}. This appears natural if one takes 
into account that the hard quarks producing the weak bosons have 
a narrower transverse spatial distribution than the soft partons.
As a result, the average impact parameters in events with weak boson
production are much smaller than in generic inelastic collisions, 
leading to an enhancement of multiple soft and semi-hard 
interactions \cite{Frankfurt:2003td}. 

Single parton densities and GPDs do not carry information about
longitudinal and transverse correlations of partons in the hadron wave
function. Such information can be extracted from high energy $pp$ and
$pA$ collisions where two (or more) pairs of partons can collide to
produce multiple dijets, with a kinematics distinguishable from those
produced in $2 \rightarrow 4$ parton processes. Since the momentum scale 
of the hard interaction, $p_t$, corresponds to much smaller transverse 
distances in coordinate space than the hadronic radius, in a double
parton collision the two interaction regions are well separated in
transverse space. Experimentally, one measures the ratio
\begin{equation}
{{d\sigma(p+\bar p\to jet_1+jet_2 +jet_3+\gamma)\over d\Omega_{1,2,3,4}} \over {{d\sigma(p+\bar p\to jet_1+jet_2 )\over d\Omega_{1,2}} \cdot  {d\sigma(p+\bar p\to jet_3+\gamma)\over d\Omega_{3,4}}}} = {f(x_1,x_3)f(x_2,x_4)\over \sigma_{eff} f(x_1)f(x_2)f(x_3)f(x_4)},
\end{equation}
where $f(x_1,x_3), f(x_2,x_4)$ are the longitudinal light-cone double parton 
densities at the hard scale $\mu^2$ (we assume for simplicity that 
the virtuality in both hard processes is comparable; in the following 
equations we suppress dependence on $\mu^2$), and the quantity 
$\sigma_{\rm eff}$ can be interpreted as the ``transverse correlation area''. 
The variables $\Omega_i$ characterize phase volume of  the observed jets (or photons).

Parton correlations can emerge due to nonperturbative effects
at a low resolution scale, or due to the effects of QCD evolution. 
One possible nonperturbative mechanism is the existence of
``constituent quarks'' within the nucleon, which appear due
to the interaction of current quarks with localized non-perturbative 
gluon fields, resulting in local short--range correlations 
in the transverse spatial distribution of gluons. The
instanton model of the QCD vacuum suggests a constituent quark
radius of about $1/3$ the nucleon radius, $r_q \approx  r_N / 3$.
Another nonperturbative mechanism, relevant at small $x$, are 
fluctuations of the color field in the nucleon due to the fluctuations 
of the transverse size of the quark distribution.
Perturbative correlations emerge due to small transverse distances
in the emission process in the perturbative partonic ladder in
DGLAP evolution. Of all the mentioned mechanisms, only the first one 
is effective at $x\ge 0.05$, where the data on production 
of two balanced jets, and jet plus photon
\cite{CDF} 
were collected. They  found
$\sigma_{\rm eff} = 14.5\pm 1.7^{+ 1.7}_{-2.3} \; {\rm mb}$ which  is significantly smaller than the naive estimate
obtained by taking a uniform distribution of partons of a
transverse size determined by the e.m.\ form factor of the nucleon,
which gives $\sigma_{\rm eff} \approx 53 \; {\rm mb}$, indicating 
strong correlations between the transverse positions of partons 
in the transverse plane. The longitudinal correlation between partons 
in the measured kinematics due to energy conservation is likely to be small,
as $x_1 + x_2$ and $x_3 + x_4$ are much smaller than 1. If this effect 
were important it would likely lead to a suppression of the double 
parton collision cross section, and hence to an increase
of $\sigma_{\rm eff}$. However, no dependence of $\sigma_{\rm eff}$ 
on $x_i$ was observed in the experiment.

We can calculate $\sigma_{eff}$ using
the information about the transverse spatial distribution of
gluons gained from $J/\psi$ photoproduction, as summarized
above. Since the $x$ values of the partons 
probed were reasonably small compared to 1, the simple ``geometric''
picture of the $\bar p p$ collision in transverse position 
in the spirit of Eq.~(\ref{P_2}) is justified,
and one has
\begin{equation}
\sigma_{\rm eff} \;\; = \;\; 
\left[ \int d^2 b \; P_2^2(b) \right]^{-1} .
\end{equation}
Evaluating this with the dipole parametrization of the 
two--gluon form factor (\ref{dipole}), this comes to $\sigma_{\rm eff} \;\; = \;\; 
28\pi /m_g^2
\;\; \approx \;\; 34\; \mbox{mb} $. 
%\end{equation}
Thus, about 50\% of the enhancement compared to the naive estimate
of the previous paragraph is due to smaller actual transverse radius
of the gluon distribution. Still, our value indicates significant
correlations in the transverse positions of the partons.
In the kinematics discussed here the relevant partons are both 
quarks and gluons. We can estimate the effect of correlations
assuming that most of the partons are concentrated in a small 
transverse area associated with the ``constituent quarks'',
as implied by the instanton liquid model \cite{Diakonov:2002fq}. Assuming a constituent quark
radius of $r_q \sim r_N/3$, we obtain an enhancement factor
due to transverse spatial correlations of partons of
$\frac{8}{9} + \frac{1}{9}\; \frac{r_N^2}{r_q^2} \;\; \sim \;\; 
1.6 \div 2$ .
%\end{equation}
This is roughly the value needed to explain the remaining discrepancy
with the CDF data. Thus, the combination of the relatively small 
transverse size of the distribution of large--$x$ gluons
and the quark--gluon correlations implied by ``constituent quarks''
with $r_q \approx r_N / 3$ is sufficient to explain the trend of 
the CDF data. Further studies of multijet events at hadron--hadron colliders, 
with a broader range of final states, would in principle allow to measure 
separately quark--quark, quark--gluon, and gluon--gluon correlations 
for different $x$.
 
However, studies based on $\bar pp$ or $pp$ collisions alone
do not allow for a model--independent separation of transverse and 
longitudinal correlations. This is possible only in $pA$ collisions
at RHIC and LHC.  The reason is that the nucleus, having a thickness 
which practically does not change on the nucleon transverse scale, 
provides an important contribution which is sensitive only to the 
longitudinal correlations of hadrons \cite{Strikman:2001gz}.
This is the contribution when two partons of the incident nucleon 
interact with partons belonging to two different nucleons in the nucleus, 
$\sigma_2$,
\begin{equation} 
\sigma_2 \;\; =\;\; \sigma_{\rm double}^{NN} \frac{A-1}{A}
\int d^2b \; T^2(b) \; \frac{f(x_1) f(x_2)}{f(x_1,x_2)} .
\end{equation}
The other term is the impulse approximation --- two partons of
the incoming nucleon interact with two partons of the same 
nucleon in the nucleus, $\sigma_1$, which is simply equal to 
$A$ times the cross section of double scattering in $pp$ collisions.
Thus, by measuring the ratio of $pA$ and $pp$ double scattering cross
sections we can determine $1/\sigma_{\rm eff}$ . 
Taking the CDF value of $\sigma_{\rm eff}\sim 14 \; {\rm mb}$, we obtain
$\sigma_2/\sigma_1\sim 3$ for $A\sim 200$. Thus, the separation of the two terms will be 
quite straightforward. Even in the case of deuteron--nucleus scattering, 
which was studied at RHIC recently, the contributions from two partons 
of one nucleon of the deuteron interacting with two different nucleons 
in the nucleus remains significant.It constitutes about 50\% of the 
cross section for $A \sim 200$. Hence it will be possible to measure $\sigma_{\rm eff}$  in  $pA$ and $dA$ collisions  if it is $\ge 5 \; {\rm mb}$, with $pA$ being 
a better option. Finally, if $\sigma_{\rm eff}$ will have been 
measured in $pA$ collisions, it will be possible to extract the
longitudinal two--parton distributions in a model independent way.

To summarize, we have demonstrated that future experiments will 
be able to measure independently the longitudinal and transverse 
two--parton distributions in the nucleon. With a detector of 
sufficiently large acceptance it would be possible to extend 
these studies even to the case of three parton correlations.

\section{Fixed target opportunities}

Experiments at fixed target facilities: BNL, CERN,  FNAL, Serpukhov were performed  for many years. However  most of the these experiments were planned in the pre QCD/ early  QCD period.
From the angle of current studies in QCD  the most  interesting topic is   large momentum transfer semi/exclusive processes with small cross sections  which were beyond  capabilities of previous machines - beam intensity, detector electronics. Here I discuss briefly several possible directions: study of quark exchanges in pQCD via two body processes, large angle two body processes, color transparency phenomena, branching $2\to 3$ processes. I will not have time/space to cover very promising direction of  the  study of the properties on cold dense nuclear matter - structure of the short-range correlations using large momentum exclusive processes with high-energy hadron beams.

\subsection{ Quark exchanges in pQCD via two body processes}

In the Regge theory two body processes in the limit of fixed t and $s\to \infty$ is given by the exchange of the corresponding Regge trajectory, $\alpha_R(t)$ with the amplitude given by $A(s,t)= f(t) s^{\alpha_R(t)}$. This amplitude is usually assumed to be linear function of t. Two simplest types of processes with largest cross sections (except the ones where vacuum Pomeron exchange is allowed) are $\pi^-p\to \pi^0 n$ where exchange by a $q\bar q$ is allowed and  $\pi^-p\to p  \pi^-$ where exchange by three quarks is allowed. In pQCD we may expect that  for large t instead of exchange by a meson/ baryon trajectory an exchange by (anti) quarks should dominate which are reggeized  in pQCD \cite{FadinS}.
Since at large t the QCD coupling is a weak function of t, $\alpha_q(t)$ should weakly depend on t.
Hence using Azimov displacement relation \cite{Azimov} we find
\begin{equation}
A_{q\bar q} \propto s^{2\alpha_q(t) -1}, \, A_{qqq} \propto s^{3\alpha_q(t) -2},
\end{equation}
leading to a an expectation of nearly t-independent meson and baryon trajectories at at large negative t with 
\begin{equation}
\alpha_N(t) =3\alpha_M(t)/2 -0.5 \approx const.
\end{equation}
We inspected the current data which are very fragmentary and found they are consistent with
$\alpha_M(-t \> 1\, GeV^2) = - (0.2\div 0.4), \alpha_B(-t \> 1\, GeV^2) = - (0.8\div 1.1)$ which is consistent with $\alpha_q(-t \> 1\, GeV^2) \sim 0.3\div 0.4$  as compared to the case of nonreggeized quark exchange of $\alpha_q=0.5$ indicating that pQCD sets up already at $-t \sim 1 GeV^2$ and that reggeization effect is rather small. One may expect that at  large t the coupling of $q\bar q$ to the meson and baryon vertices is given by the corresponding GPDs.
Note also that in the case of baryon exchange the pQCD diagrams involve $q\bar q g$ configurations in mesons, cf \cite{FS81}.
\subsection{Large angle two body processes}
So far we do not understand the origin of one of the most fundamental hadronic processes in pQCD -large angle two body reactions 
($-t/s=const,  s\to \infty$): $\pi +p \to \pi +p, p+p\to p+p$. The most extensive set of processes was studied in the BNL experiment at 5.9 and 9.9 GeV/c \cite{White:1994tj}. The data indicate dominance of the quark exchanges and appear to be consistent with quark counting rules\cite{Brodsky:1973kr} based dominance of the minimal Fock components in the wave functions of the colliding hadrons. 

We find that several features of the data are consistent with dominance of small size configurations. In particular
\begin{equation}
{d\sigma^{K^+p\to K^+p}\over d\theta_{c.m.}}(\theta=90^o) /
{d\sigma^{\pi^+p\to \pi^+p}\over d\theta_{c.m.}}(\theta=90^o) \sim (f_K/f_{\pi})^2\sim 1.7,
\end{equation}
while the data gives $1.69 \pm 0.25$. Also  
\begin{equation}
{d\sigma^{\pi^+p\to \pi^+p}\over d\theta_{c.m.}}(\theta=90^o) / 
{d\sigma^{\pi^-p\to \pi^-p}\over d\theta_{c.m.}}(\theta=90^o)\sim u(x)/d(x)\sim 2, 
\end{equation}
agrees with the data with accuracy $10\div 15\%$ both for elastic and for $\rho$-meson production channel. Overall it appears likely that these processes are dominated by short distances for -$t> 5 \mbox{GeV}^2$.  Clearly new experiments are necessary to determine details of the dynamics. J-PARC is in the optimal energy range. 
\subsection{Color transparency phenomena}
At high energies weakness of  interaction of point-like configurations with nucleons - is routinely used for explanation of DIS phenomena at  HERA.

First experimental observation of high energy weakness of interaction with nuclei - color transparency (CT)  - for pion interaction was reported for the process:
$\pi  +A \to ÓjetÓ+ÓjetÓ +A$ in \cite{Ashery}. It confirmed our predictions of pQCD  for A-dependence, distribution over energy fraction, $u$ carried by one jet, dependence on $p_t(jet)$, etc.
There is also experimental evidence for CT phenomenon in exclusive produciton of $\rho$ and $\pi$- mesons by virtual photons - see \cite{Dutta} and references therein.

Main issues are (a) 
at what $Q^2  (t)$   particular processes start to be dominated by  point-like configurations   -  for example interplay of end point and LT contributions in the e.m. form factors, (b) If the point-like configuration is formed,   how long it will remain  smaller than average configuration? 
This involves both expansion after interaction to an average configurations and contraction before interaction  from an average configurations. The recent data\cite{Dutta}  see to be in agreement with an early estimate of this phenomenon within the color diffusion model \cite{FFLS88}. In particular, the data are consistent with the estimate of \cite{FFLS88} of the coherence length $l_{coh}=(0.3 \div 0.4 ) fm\cdot  p_h/GeV$. (It is worth noting that this coherence length is much smaller than the one usually assumed in the heavy ion collisions).

In spite  of progress with studies of CT with virtual photons, investigation of  CT for the hadronic projectiles remains a  challenge - no definitive conclusions were reached in the BNL experiments. It is critical to perform new studies of CT phenomenon in hadronic reactions at energies above 10 GeV where expansion effects are moderate.  Such measurements would  complement the program of CT in eA scattering at Jlab at 12 GeV.
Natural places to perform such experiments would be J-PARC and FAIR at GSI. Obvious 
advantages  as compared to the previous round of experiments are progress in electronics leading to a possibility to work at higher luminosity, wider range of hadron beams including antiprotons at GSI,  possibility  of using polarized beams. It would be  desirable to perform (p,2p) experiments at the range of 10-20 GeV for all angles including 
those close to $\theta_{c.m.} \sim 90^0$ and extend measurements to 
$E_p >20 GeV$. In the latter case the rates at  $\theta_{c.m.} \sim 90^0$ 
are too small, and one should probably adopt another strategy - measure transparency for  large but fixed t.  In this case $l_{coh}$ for initial and the fastest of two final nucleons is very large. Only the slow nucleon has time to expand leading to a large CT effect --transparency  similar to the one in $A(e,eÕp)$ reaction. 

\subsection{Branching processes and GPDs}
Detectors which can study CT are well suited also to  study generalized parton distributions using hadronic projectiles complementing the studies with lepton projectiles. Such studies  will be especially beneficial if they will be performed   in parallel with 12 GeV program at Jlab (GPD studies is the main trust of Jlab  12 GeV program).

The idea is to consider new type of hard hadronic processes - branching exclusive  processes of large c.m.angle scattering on a ÒclusterÓ in a target/projectile 
\cite{baryon95}. Examples of such reactions include 
$h +p \rightarrow h' + M +B$ where M is a meson carrying
most of the transfered momentum, while $B$ is recoil system which
 mass and momentum are kept fixed in the target rest frame, and  
 reverse processes in which  a fast baryon is ejected from
the target: 
$h +p \rightarrow h' + B +M$ (Fig.1).
 Among interesting channels are production of exotic meson states in the recoil kinematics, study of hidden/intrinsic strangeness and  charm in hadrons in the processes like $pp \to \Lambda(spectator) + K^+ + p, \Lambda + K^+(spectator) + p, \phi(spectator) + p +p$. It would be also advantageous to use polarized beams and/or   targets. 
 
 In the limit when CT is valid for the elementary $2\to 2$ reaction one can try to connect cross sections of processes where $q\bar q$ pair scatters at large angles to the nucleon GPDs in the kinematics where x's of both removed partons are positive (ERBL region). 
 Similarly in the process $pp \to N + \pi +\Delta (spectator)$ one would probe non-diagonal  $N\to\Delta $ GPDs. Theoretical calculations along these lines  are under way  \cite{Kumano}.
 
 To summarize,
 in order to understand the complexity of the  structure 
of  nucleons (as well as other hadrons) a diverse program 
 studies of parton correlations, fluctuations of the size of the hadrons, structure of minimal and higher order Fock components of the hadron wave function are necessary. This would require use of both colliders  
and fixed target facilities, combining leptonic and hadronic probes. The currently opertating facilities and facilities which are now under construction provide excellent  opportunities for such studies. 
                         
                         \begin{figure}
\includegraphics[width=1.0\columnwidth]{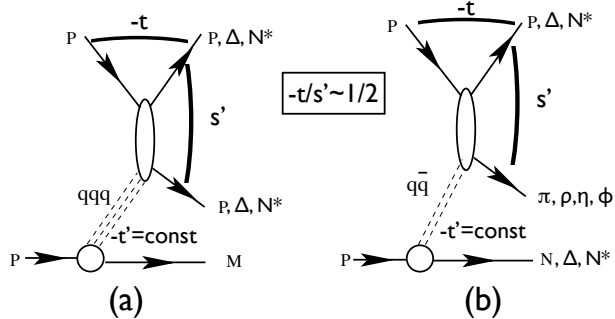}
\vspace{-3.2cm}
\caption{\it (a) Production of fast baryon and recoiling mesonic system, (b) Production of fast meson  and recoiling baryonic system.}
\end{figure}

\end{document}